\documentclass[aps, twocolumn,floatfix,showpacs,preprintnumbers,superscriptaddress,amsmath,amssymb,pre]{revtex4-1}

\usepackage{xcolor}
\usepackage{color}

\usepackage{wasysym}
\usepackage{mathcomp}
\usepackage{graphicx}
\usepackage{dcolumn}
\usepackage{bm}
\usepackage{hyperref}
\usepackage[mathlines]{lineno}
\usepackage{comment}
\includecomment{pablocomments}
\usepackage{tikz}
\usepackage{pgfplots}
\usetikzlibrary{shapes}

\usepackage[caption=false]{subfig}

\usepackage{lipsum}  

\newcommand*{\tikzbullet}[2]{%
  \setbox0=\hbox{\strut}%
  \begin{tikzpicture}
    \filldraw[draw=#1,fill=#2] (0,3\ht0) circle[radius=.25em];
  \end{tikzpicture}%
}

\newcommand*{\tikzrectangle}[2]{%
  \setbox0=\hbox{\strut}%
  \begin{tikzpicture}
    \filldraw[draw=#1,fill=#2] (0,3\ht0) rectangle ++(5pt,5pt);
  \end{tikzpicture}%
}

\newcommand*{\tikztriangle}[2]{%
  \setbox0=\hbox{\strut}%
  \begin{tikzpicture}
    \node[draw=#1,fill=#2,regular polygon, regular polygon sides=3,inner  sep=1.3pt] at (5cm,0) {};
  \end{tikzpicture}%
}

\newcommand*{\tikzline}[1]{%
  \setbox0=\hbox{\strut}%
  \begin{tikzpicture}
    \useasboundingbox (-0.1em,-0.2em) rectangle (1.5em,\ht0);
    \draw[color=#1,solid,line width=0.5pt](0,0) -- (5mm,0);
  \end{tikzpicture}%
}

\newcommand*{\tikzdashedline}[1]{%
  \setbox0=\hbox{\strut}%
  \begin{tikzpicture}
    \useasboundingbox (-0.1em,-0.2em) rectangle (1.5em,\ht0);
    \draw[color=#1,dashed,line width=0.5pt](0,0) -- (5mm,0) {};
  \end{tikzpicture}%
}

\newcommand*{\tikzdottedline}[1]{%
  \setbox0=\hbox{\strut}%
  \begin{tikzpicture}
    \useasboundingbox (-0.1em,-0.2em) rectangle (1.5em,\ht0);
    \draw[color=#1,dotted,line width=0.5pt](0,0) -- (5mm,0) {};
  \end{tikzpicture}%
}

\newcommand*{\tikzdashdotline}[1]{%
  \setbox0=\hbox{\strut}%
  \begin{tikzpicture}
    \useasboundingbox (-0.1em,-0.2em) rectangle (1.5em,\ht0);
    \draw[color=#1,dash dot,line width=0.5pt](0,0) -- (5mm,0) {};
  \end{tikzpicture}%
}

\begin{document}

\title{Micro-mechanical response and power-law exponents from the longitudinal fluctuations of F-actin solutions.}

\author{Pablo Dom\'{i}nguez-Garc\'{i}a}
\affiliation{Dep. F\'{i}sica Interdisciplinar, Universidad Nacional de Educaci\'{o}n a Distancia (UNED), Madrid 28040, Spain.}
\author{Jose R. Pinto} 
\affiliation{Department of Biomedical Sciences, Florida State University College of Medicine, Florida, USA. }
\author{Ana Akrap}%
\affiliation{Department of Physics, University of Fribourg, Fribourg, Switzerland.}

\author{Sylvia Jeney}%
\affiliation{Department of Physics, University of Fribourg, Fribourg, Switzerland.}

\begin{abstract}
We investigate the local fluctuations of filamentous actin (F-actin), with focus on the skeletal thin filament, using single-particle optical trapping interferometry. 
This experimental technique
allows us to detect the Brownian motion of a tracer bead immersed in a complex fluid with nanometric resolution at the microsecond time-scale.
The mean square displacement, loss modulus, and velocity autocorrelation function (VAF) of the trapped microprobes in the fluid follow power-law behaviors, whose exponents can be determined in the short-time/high-frequency regime along several decades. 
We obtain 7/8 subdiffusive power-law exponents for polystyrene depleted microtracers at low optical trapping forces. 
Microrheologically, the elastic modulus of these suspensions is observed to be constant up to the limit of high frequencies, confirming the origin of this subdiffusive exponent on the local longitudinal fluctuations of the polymers.
Deviations from this value are measured and discussed in relation to the characteristic lengths scales of these F-actin networks and probes' properties, and also in connection with the different power-law exponents detected in the VAFs.   
Finally, we observe that the thin filament, composed by tropomyosin (Tm) and troponin (Tn) coupled to F-actin in the presence of Ca$^{2+}$, returns exponent values less dispersed than F-actin alone, which we interpret as a micro-measurement of the filament stabilization.

\end{abstract}

\maketitle
\section{Introduction}

Optical tweezers (OT) have been demonstrated as a revolutionary technique in soft matter, complex fluids and biophysics since Ashkin and co-workers published their seminal paper \cite{Ashkin1986}. Using focused beams of laser light, OT allow to trap and control tracer micrometer-sized particles suspended in a complex fluid, and, therefore, the study of such fluid
on biologically relevant scales. 
Since the position fluctuations of the trapped tracer depend on the properties of the surrounding medium, 
the viscoelastic properties of the fluid can be inferred by microrheology \cite{mason_optical_1995}. However, even very small forces applied to the fluid (usually a polymeric suspension) may affect the mechanical properties of the bead's surroundings \cite{Tassieri2015, Dominguez2016}. 
This drawback becomes interesting when we focus on the interaction among the colloidal particle and the polymers which defines the fluid behavior. 
Besides, the optical force may affect the colloid-polymer system, already complex because of the interaction with the biomaterial, the compressibility of the matrix network, and the polymer adsorption and depletion effects near the colloid surface \cite{Donath1997, chen_rheological_2003, valentine_colloid_2004, He2011}. 
In practice, the understanding of colloid-polymer interactions is important for their role in industrial applications, from food technologies to personal care products \cite{larson_structure_1999}.  

The motion of single-tracers is then used to deduce the mechanical properties of the fluid. One-particle (1P) microrheology provides the fluid viscoelasticity on the microscale, similar to the tracer's size, i.e., microrheology depends on the length scales of the system \cite{Atakhorrami2014}, as opposed to bulk rheology \cite{Gardel_microrheology_2003}. 
To overcome the limitations of 1P microrheology, two-particle (2P) microrheology \cite{levine_one-_2000, crocker_two-point_2000} detects the fluctuations at large length scales, matching bulk rheology as a result. 
F-actin solutions are a classic example of substantial differences between micro- and macrorheology results, but a correct characterization of the involved length scales 
clarifies some of the discrepancies in other viscoelastic fluids \cite{Buchanan2005}. 
The advantage of classical 1P microrheology relays on its potential capacity for isolating the local contribution to the fluctuations in the scale of the tracer bead, therefore, allows to investigate the polymer dynamics in the biological scale.

As a model case of semiflexible polymer, we study here suspensions of filamentous actin (F-actin). 
This polymer is the most abundant protein of the cytoskeleton in eukaryotic cells and determines their mechanical properties \cite{Bao2003}, being also involved in different bioprocesses, remarkably muscle contraction \cite{Bershitsky1997}. 
We focus on the local mechanical properties of F-actin when adding skeletal muscle tropomyosin (Tm) alone, and Tm and troponin (Tn). 
Tm is a coiled-coil protein, which, together with actin and Tn, constitutes the thin filament. 
The coupling of Tm/Tn complex to actin, regulated by the presence of Ca$^{2+}$, allows the interaction between myosin and actin \cite{Ebashi1971}, and thus muscle contraction. Mechanically, Tm/Tn complex tends to stabilize the filament structure and increases its stiffness, but without major modifications in the network structure \cite{Gotter1996}.

In this work, we study the power-law behavior of F-actin solutions, with the aim of rationalizing the variability of the reported subdiffusive exponents which deviates from the expected results for semiflexible polymers.
Since F-actin solutions have been object of intense investigations during the past three decades in the context of biophysics and microrheology, we will focus here on the skeletal thin filament, since we will obtain that the F-actin/Tm/Tn complex has a very similar micro-mechanical response to simple F-actin networks.
The exponents here calculated are related to the local actin longitudinal fluctuations, which are dependent on the colloid size/polymer length scales, and on the external forces/bead's surface characteristics. 
The inspection of the motion of optically trapped microtracer beads inside these polymeric networks may provide an insight of the complex colloid-polymer interplay, specially in the presence of external optical forces. 
The understanding of the colloid-polymer interaction under the influence of external forces and the mechanical response of these biopolymeric networks in the microscale may expand the knowledge of the microscopical origins of their rheology and our capacity to emulate biological systems.

\section{Experimental methodology}

We measure the motion of single optically-trapped tracer beads using optical trapping interferometry (OTI), composed by an optical trap combined with an interferometric position detector, 
allowing measurements of the Brownian position fluctuations of the microbead with nanometric resolution in the microsecond range \cite{franosch_resonances_2011}. 
This OTI system is based on a Nd:YAG laser, with a wavelength of $1064\,$nm and maximum power of $2\,$W (MEPHISTO Innolight, GmbH). 
The optical trap is formed by a 10x beam expander and a 60x water immersion objective (Olympus UPlanApo/IR, NA = $1.2$). The trajectory of the trapped tracer is recorded through the back focal plane detection by the interference between forward-scattered light from the bead and unscattered light. This signal is monitored using a quadrant photodiode positioned on the optical axis at a plane conjugate to the back focal plane of the condenser. 
Regarding heating effects through laser absorption, the maximal laser power used in these experiments is 200 mW at the focus, which corresponds to an increase of less than $2\,^\circ$C in water \cite{Petterman2003}. 

We disperse $1\,\mu$l of microspheres suspension in $200\,\mu$l of working suspension (described later), 
which is loaded into the interior of a rectangular custom-made flow cell. 
The low concentration of beads ensures that the distance between particles is high enough to minimize their interactions. 
The flow-cell is composed by two pieces of double-sided tape with size $2\,\times 0.5\,$cm and thickness $100\,\mu$m,  
and mounted upside down on the 3D piezo-stage \cite{lukic_direct_2005}. 
Using the piezo stage, the vertical position of the optically trapped bead is displaced to the center of the container ($z=50\,\mu$m), 
such that the top and bottom glass surfaces have no influence in the microsphere motion \cite{faucheux_confined_1994,jeney_anisotropic_2008}. 
The motion of each tracer is collected during $100\,$s at a sampling rate of $1\,$MHz, generating $10^8$ data points per measurement. 

In these experiments, we used two types of monodisperse microbeads as tracers: melamine resin microbeads (Microparticles, GmbH) with radius $a=1.47\,\mu\text{m}$ and density $\rho_p = 1570\,$kg/m$^3$ at $T= 21^\circ$C, and polystyrene (PS) particles (Sigma-Aldrich) with $a=0.99\,\mu\text{m}$ and $\rho_p = 1065\,$kg/m$^3$ ($T= 21^\circ$C). 
Particularly important for this study is the colloid surface chemistry of the microbeads \cite{valentine_colloid_2004}.
Polystyrene latex is hydrophobic, generating the adsorption of polyelectrolytes and proteins when it is immersed in an aqueous solution. To reduce the polimeric binding to the particle surface of the PS particles, we mixed these beads with adsorbed bovine serum albumin (BSA, Sigma-Aldrich). BSA is usually preadsorbed to probes in order to block the surface and reduce their interaction with F-actin filaments \cite{McGrath2000}. The PS beads are incubated overnight in a $200\,$mg/ml BSA solution, and prior to their use, washed in G-buffer with several successive centrifugation and redispersion steps to remove unbound BSA or impurities \cite{Chae2005}.
The melanine resin beads are composed of a thermosetting plastic which has been proved to remain stable in series of organic compounds \cite{Meyer2006}, and they have been found to be more representative than PS particles of the actual microstructure of the network \cite{Samiul2012}.

Actin has a monomeric globular form (G-actin) at low ionic strength, with an approximate diameter of $5\,$nm. G-actin polymerizes to actin filaments (F-actin) when the ionic strength is increased to physiological values. 
F-actin has a diameter of $d\sim 7\,$nm \cite{Egelman1985} and an average contour length $l \sim 20\,\mu$m \cite{Kaufmann1992}, persistence length $l_p \sim 17\,\mu$m \cite{Ott1993} (in the range $9-20\,\mu$m depending on preparation \cite{Isambert1995}), with mesh size \cite{Schmidt1989} $\xi = 0.3/\sqrt{c_A}$, expressed in $\mu$m if $c_A$ is the actin concentration in mg/ml.

Actin powder was purchased from Cytoskeleton, Inc. (catalog AKL99), which was purified from rabbit skeletal muscle using the well-established method of Pardee and Spudich \cite{Pardee1982} and has a level of purity greater than 99\%. The F-actin was polymerized following the protocol for storage and reconstitution of the lyophilized protein provided by Cytoskeleton, Inc.  A brief centrifugation step of at $16\,000 \times g$ for 15 minutes at a temperature of 4$^\circ$C was introduced during the polymerization process to remove non-polymerizable actin and to reduce the presence of nucleation sites.
The lyophilized monomeric actin was reconstituted by dissolving $1\,$mg in $100\,\mu$l
of Milli-Q water, which gives a concentration of $10\,$mg/ml in the following G-buffer: $5\,$mM Tris-HCl pH $8.0$, $0.2\,$mM CaCl$_2$, $0.2\,$mM ATP (Adenosine triphosphate), 5\% (wt/vol) sucrose, and 1\% (wt/vol) dextran. 
The actin is diluted and polymerized into a concentration of $1\,$mg/ml ($23.8\,\mu$M) F-actin by mixing $500\,\mu$l of G-actin with $55\,\mu$l of the following (10x solution) buffer: $10\,$mM Hepes pH $7.5$, $0.2\,$mM CaCl$_2$, $0.2\,$mM ATP, $50\,$mM KCl, $2\,$mM MgCl$_2$, $1\,$mM dithiothreitol (DTT).
We also added $14.9\,\mu$l of Phalloidin ($1\,$mM) in a 1:1 molar ratio to stabilize the structure of actin filaments \cite{Wulf1979}. 

The thin filament, composed by tropomyosin (Tm) and troponin (Tn) coupled to actin, and regulated by the presence of calcium \cite{Ebashi1971}, has a role in its interaction with myosin regarding muscle contraction. Tm is an extended molecule with length of $\sim 42\,$nm composed by a dimer of two polypeptide subunits. 
This molecule winds around the actin filament and is able to attach to other tropomyosin molecules into long strands, spanning the whole thin filament length \cite{Khaitlina2015}. Troponin is composed by three subunits: TnT, TnI and TnC. The subunit TnI interacts with actin, producing changes that prevent the interaction with myosin. 
Tm relates to TnI by facilitating the transmission of the changes on actin generated by TnI peptide along the filament \cite{Perry2003}. 
The calcium-regulation of the myosin-induced contraction depends on the interaction between the actin filaments, TnC and TnI \cite{Potter1975}. Mechanically speaking, the Tm/Tn complex stabilizes the F-actin structure by increasing the stiffness of the compound \cite{Gotter1996}.

Here, skeletal muscle tropomyosin and troponin \cite{Ebashi1971} were purified from rabbit muscle, according to standard methodologies \cite{Potter1982}. Tn, Tm, and F-actin in G-buffer were mixed at a 1:1:7 stoichiometry ($3.31\,\mu$M Tn, $3.31\,\mu$M Tm, and $23.17\,\mu$M F-actin) according to the molar proportion in striaded muscle \cite{Irving2012}. Before its use, the protein mixture was incubated for an hour at 4$^\circ$C. We note that, according to the previously described preparation of actin buffer, all experiments were performed in the presence of Ca$^{2+}$ ($0.2\,$mM CaCl$_2$).

\section{Results}

\begin{figure}
\centering
\includegraphics[scale=1]{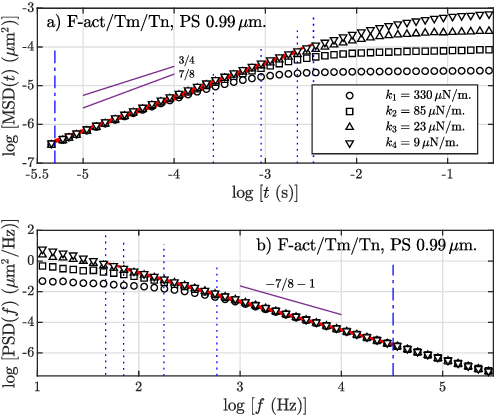}
\caption{\label{fig:sup8} a) One dimensional particle mean-squared displacement (MSD) 
and b) Power spectral density (PSD) for $1\,$mg/ml F-actin solution with Tm and Tn for an optically trapped 
polystyrene (PS) BSA microparticle with $a=0.99\,\mu$m. 
The data have been blocked in 10 bins per decade. Errors are very small and are not plotted for clarity. Trap forces are indicated in the legend. 
Regression lines (\protect\tikzline{red} ) are calculated under the limits defined by the dashed lines, the purple line (\protect\tikzline{purple} ) is a guide for the eye indicating the expected power-law exponents values.}
\end{figure} 

The most applied correlation quantity in microrheology is the mean square displacement (MSD) of the location, $x(t)$, of the probe tracers, defined as $\textrm{MSD}(t) \equiv \left<( x(t)-x(0))^2\right>$ (in one dimension), along with the power-spectral density (PSD), which is calculated by Fourier modes $x_T(f)=\int^{T/2}_{-T/2}\exp(ift/2\pi)x(t)dt$ and then applying
$\textrm{PSD}(f) \equiv \lim_{T \rightarrow \infty} (1/T)\left<\left|x_T(f)\right|\right>$. 
For an optically trapped particle in a Newtonian fluid,  when the optical trap is predominant at large lag times, the MSD reaches a plateau $\textrm{MSD}(\infty)=2k_B T/k$, where $k$ is the stiffness of the optical trap, following the equipartition theorem. At intermediate and low time values, it follows a power-law $\textrm{MSD}(t) \sim t^{\alpha}$, with $\alpha = 1$ defining the diffusive behavior, before the influence of hydrodynamic effects, where the ballistic region appears \cite{huang_r_direct_2011}. 
For many complex fluids, the power-law behavior also shows up, but with a subdiffusive exponent $\alpha<1$. 
The PSD of a microbead in a Newtonian fluid has a Lorentzian form, which can be approximated to PSD$(f) \sim f^{-2}$ at high frequencies, but where no hydrodynamics are considered. It is possible to observe deviations from the Lorentzian spectra reflecting the color of thermal noise \cite{franosch_resonances_2011}. A generic power-law fluid with MSD$(t)\sim t^{\alpha}$ should show a PSD$(f)\sim f^{-\alpha-1}$ power-law behavior.

In Fig. \ref{fig:sup8}, we show an example of MSD and PSD curves for a micron-sized PS tracer bead optically-trapped with different trap stiffnesses and immersed in a solution of the F-actin/Tm/Tn complex. The dashed lines are the limits of the linear regressions, which are shown in red color (see supplementary material about the criterion to assign these limits). The MSD reaches a plateau at higher times because of elastic forces from the trap and the fluid itself. At intermediate and short-time values, it verifies an subdiffusive power-law behavior. 
The behavior of the PSD is analogous, but in the frequency space. 
Apparently, the power-law behaviors for PS BSA-added depleted particles generate a $\alpha \approx 7/8$ value from both statistical quantities. Similar curves are obtained from the different combinations in F-actin solutions (F-actin alone and F-actin with Tm), 
and using resin tracers, with no appreciable differences between them (see the supplementary material for a complete set of figures). 

\begin{figure}
\centering
\includegraphics[scale=1]{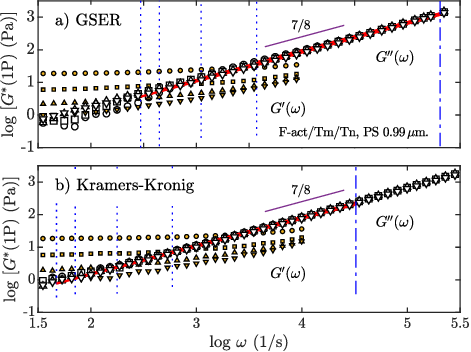}
\caption{\label{fig:sup9} Storage modulus, $G'(\omega)$ (orange-filled symbols) and loss modulus, $G''(\omega)$ (white-filled symbols), for an optically trapped polystyrene (PS) BSA microparticle with $a=0.99\,\mu$m in F-actin/Tm/Tn complex ($c_A = 1\,$mg/ml). 
a) Obtained by means of the GSER and the MSD, or b) through the Kramers-Kronig expression and the PSD. The data symbols are the same as in Fig. \ref{fig:sup8}. Regression lines are calculated under the limits defined by the dashed lines. $G'(\omega)$ shows a characteristic breakup at high frequencies (see Supp. Material.)}
\end{figure}

From macro-rheometry and compatible methods, such as two-point microrheology, the observed exponent for a F-actin solution of semiflexible polymers \cite{morse_viscoelasticity_1998, gittes_dynamic_1998}, like the one used in these experiments, should be $\alpha=3/4$. On the other hand, an exponent $\alpha = 7/8$ is obtained in F-actin solutions when adding myosin \cite{LeGoff2001} and also appears in some experiments similar to the reported here \cite{Chae2005,Koenderink2006,atakhorrami_short-time_2008,Tassieri2008,tassieri_microrheology_2012,Grebenkov2013, Atakhorrami2014} (see Supp. Material for a more specific enumeration). 
Apart from experiments, the exponent 7/8 also appears in the stress autocorrelation function of semiflexible polymers simulated by Brownian dynamics \cite{Dimitrakopoulos2001}, in the polymer fluctuations parallel to the local axis \cite{Liverpool2001}, in tension dynamics of extensible wormlike chains \cite{Obermayer2009}, and using a wormlike-chain model of semiflexible chains with internal friction \cite{Hiraiwa2010}. 
Theoretically, there are two competing anisotropic contributions in the fluctuations of the semiflexible polymer: parallel or perpendicular to the local orientation \cite{Everaers1999}, where the longitudinal fluctuations behave \cite{Gardel2004} as $\left<\delta r_\shortparallel\right>\sim t^{7/8}$. The 3/4 exponent appears when there is no dependency on the grade of filament entanglement in the network \cite{addas_microrheology_2004} and, at high frequencies, when the entanglements and the rigidity of the polymers only allow to move the filaments laterally \cite{palmer_diffusing_1999}. 

Fig. \ref{fig:sup9} shows two examples for the micromechanical properties for suspensions of the F-actin/Tm/Tn complex by plotting the real and imaginary parts of the complex modulus, $G^*(\omega) = G'(\omega)+iG''(\omega)$, where $G'$ is the storage or elastic modulus and $G''$ is the loss modulus. These quantities are obtained by means of standard and well-known microrheological methods, i.e., by means of the generalized Stokes-Einstein relation (GSER) concurrent with Mason's approximation \cite{mason_particle_1997,mason_estimating_2000} (Fig. \ref{fig:sup9}a)), and through the Kramers-Kronig integrals (Fig. \ref{fig:sup9}b)). The observed viscoelastic behavior is analogous to the one we have obtained for F-actin alone (see Supp. Material): at lower frequencies, a plateau modulus appears; but at high frequencies, the fluid is liquid-like because the loss modulus, $G''(\omega)$, dominates the elastic one, $G'(\omega)$. The loss modulus follows the expected power-law behavior $G''(\omega) \sim \omega^\alpha$, where the exponents $\alpha$ are calculated using the same procedure that in Fig. \ref{fig:sup8},  
with an $\alpha$ value compatible with $7/8$.  

In Fig. \ref{fig:sup9}, we observe constant values for $G'(\omega)$ for frequencies between $1-10\,$kHz even at the lowest optical force, without following the power-law tendency of the loss modulus at higher frequencies. 
The expected influence of the optical trap appears in the plateau of the MSD in Fig. \ref{fig:sup8} a), which ends before $1\,$kHz. Regarding the values of the elastic component at the lowest frequencies available, in our experiments each $G'(\omega)$ depends on the applied optical force by the influence of its elastic component $G'_k=k/6\pi a$.  With our optical tweezers, particles immersed in water with $a=0.99\,\mu$m generate an elastic response $G_k=0.5\,$Pa for the weakest optical force, and $G_k = 18\,$Pa for the strongest ($G_k = 0.25\,$Pa and $G_k=7\,$Pa, respectively, when $a=1.47\,\mu$m). 
In similar experiments \cite{Tassieri2008} for the cardiac thin filaments with calcium for $1\,$mg/ml actin concentration, it is obtained $G^0 = G'(\omega_0)< 0.1\,$Pa, where $\omega_0$ is the lowest available frequency. In general, the accepted value for F-actin alone at $1\,$mg/ml is $G^0 \sim 1\,$Pa, but with variations depending on preparation, type of polymerization and storage \cite{Xu1998}. However, many different investigations in the rheology of physiological F-actin solutions provide $G^0<1\,$Pa by means of different techniques in rheometry \cite{Hinner1998, Gisler1999_scaling, Schmidt2000, Goldmann2000, Koenderink2009} and in 1P/2P microrheology  \cite{crocker_two-point_2000, Gardel_microrheology_2003, Koenderink2006, He2011}.

Fig. \ref{fig:sup9} shows that, for all optical forces applied, the elastic behavior of the fluid remains approximately constant, even for the weakest optical forces ($G_k' \leq 0.5\,$Pa). This effect only begins to vanish at high frequencies when the elastic modulus grows and seems to collapse to a single unique curve. The observed elastic modulus in Fig. \ref{fig:sup9} is the result of the interplay of two interacting contributions, the fluid itself and the applied external force, which provokes a dynamical process affecting to the microstructure generated by the polymers. We have previously shown that the external optical force interacts with the network structure in other viscoelastic fluids, but that effect only appears in the storage modulus at the lowest frequencies available \cite{Dominguez2016}. In bulk rheology, where $G'(\omega)$ grows with frequency, the elastic plateau generated by the fluid appears only at very low frequencies \cite{Gardel_microrheology_2003}.

\begin{figure}
  \begin{center}
  \subfloat{\includegraphics[scale=1.05]{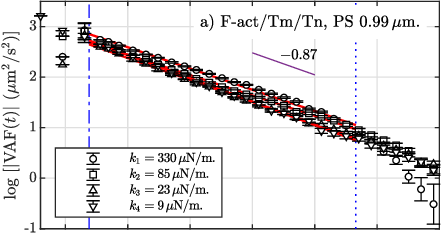}} \\
  \subfloat{\includegraphics[scale=1.05]{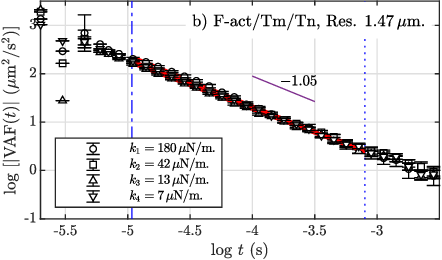}}
    \end{center}
  \caption{Absolute values of the velocity autocorrelation function, $|\text{VAF}|$ at $1\,$mg/ml F-actin with Tm and Tn solution for a) polystyrene (PS) BSA microparticle with $a=0.99\,\mu$m and b) melamine resin with $a=1.47\,\mu$m. The errors are calculated by the blocking method (see Supp. Material). Data points for the different trap stiffnesses are indicated in the legends. Regression lines (\protect\tikzline{red} ) are calculated under the limits defined by the blue dashed lines: 
top (\protect\tikzdottedline{blue} ) and lower limits (\protect\tikzdashdotline{blue} ). The purple lines (\protect\tikzline{purple} ) are a guide for the eye indicating the averaged power-law exponent $\beta$ in each case.\label{fig:sup12}} 
\end{figure}

Finally, we study the velocity autocorrelation function defined as $\textrm{VAF}(t) \equiv \left<v(t)v(0)\right>$, where $v$ is the velocity of the particle. Fig. \ref{fig:sup12}a) plots this function in absolute values, $\left|\textrm{VAF}(t)\right|$, for PS BSA-coated (a) and melanine resin (b) particles, for the available optical forces, when the fluid is composed by the F-actin/Tm/Tn complex. The VAF is related to the dissipation of the particle's motion by its interaction with the surrounding viscoelastic network. Regarding its multiple power-law behavior, it is known that, in Newtonian fluids, the VAF shows a minimum in the first zero-crossing value, related to the interplay between the optical trapping and the hydrodynamics of the fluid. Before that minimum, this function  follows an algebraic power-law rather than an exponential tail 
$\left|\text{VAF}\right|(t)\sim t^{-3/2}$, known as the ``long-time tail'' \cite{Alder1970}. A second zero-crossing appears at higher time scales in the form of $\sim t^{-7/2}$, but its experimental observation is very limited by the initial noise floor of the experimental set-up. 
In Fig. \ref{fig:sup12}, we detect a third power-law behavior, $\left|\text{VAF}\right|(t)\sim t^{-\beta}$, for an intermediate time-scale. Remarkably, this $\beta$ exponent does not depend on the applied optical forces, only on the type of tracer. 
Because no difference is observed in the obtained $\beta$ values for the same type of tracers using different fluids and optical stiffness (see averaged values in Table 1S in Supp. material), we obtain two different exponents, averaged for all $k$ values and for all combinations of F-actin solutions: $\left<\beta\right> = 0.870(6)$ for PS particles (Fig. \ref{fig:sup12}a)), and $\left<\beta\right> = 1.050(3)$ for resin tracers (Fig. \ref{fig:sup12}b)). 
In Fig. \ref{fig:sup12}a) shows a dependence of the numeric values of the VAF with the applied optical force, with the strongest optical force on top and the weakest at the bottom, but such an effect does not appear for resin tracers, as it can be seen in Fig. \ref{fig:sup12}b). The observed behavior is analogous in F-actin alone (see additional graphs in Supp. Material).

\section{Discussion}

\begin{figure}
\centering
\includegraphics[scale=1]{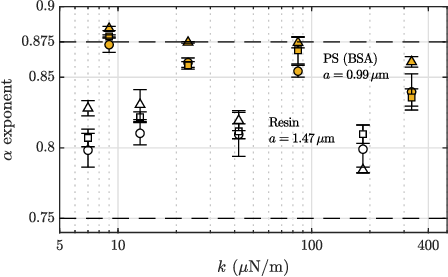}
\caption{\label{fig:fig3}Dependence of the averaged values of the local power-law exponent $\alpha$ with applied optical trap stiffness for polystyrene (PS) BSA-added (F-actin: \protect\tikzbullet{black}{orange}, F-actin/Tm: \protect\tikzrectangle{black}{orange}, and F-actin/Tm/Tn: \protect\tikztriangle{black}{orange}) and melamine resin tracers (F-actin: \protect\tikzbullet{black}{white}, F-actin/Tm: \protect\tikzrectangle{black}{white}, and F-actin/Tm/Tn: \protect\tikztriangle{black}{white}). Horizontal error bars are not included for clarity. Dashed lines (\protect\tikzdashedline{black} ) are a guide for the eye indicating exponents $\alpha=7/8$ and $3/4$. } 
\end{figure}

In this study, by analyzing the power-law behavior of the functions MSD$(t)$, PSD$(f)$ and $G''(\omega)$, we have systematically calculated $\alpha$ values for an appreciable collection of measurements, always in the short-times/high frequencies regime. 
In the case of PS particles, we obtain results compatible with an exponent $\alpha \approx 7/8$. 
However, if we focus on calculated subdiffusive exponent values under the influence of the optical forces for the two type of microtracers, we obtain Fig. \ref{fig:fig3}. 
This figure shows the averaged values for the power-law exponents using all the different F-actin solutions. 
We observe how our experiments using micro-sized PS BSA-added particles return $\alpha \simeq 0.875$ for low optical forces, but, when we increase the trap stiffnesses, the exponent slightly decreases. 
On the contrary, the exponent value for the resin particles varies around $\alpha \simeq 0.81$ with no appreciable dependency with the optical trap strength.  

As previously mentioned, the origin of the 7/8 exponent value relies on the longitudinal response provoked by a local microperturbation in the polymer generated by the tracers. The finite propagation of the tension along the filament generates a more rapid variation in the mean-squared displacement through a $t^{7/8}$ growth at low time-scales \cite{Everaers1999}. 
One-particle microrheology allows to retrieve the local fluctuations of the filaments, thus giving this exponent as a result. 
However, the micro-mechanical response for these kind of solutions depends on the type of tracer and on their surface characteristics. 
Specifically, we have to take into account the relative size of the optically-trapped tracers in relation with the characteristic lengths of the polymer network. 
Regarding the influence of skeletal Tm/Tn, this complex modifies the persistence length of actin filaments depending on the presence of calcium ($l_p =12-20\,\mu$m) \cite{Isambert1995}, but does not affect the network structure generated by F-actin \cite{Gotter1996}.

\begin{figure*}
\centering
\includegraphics[scale=1]{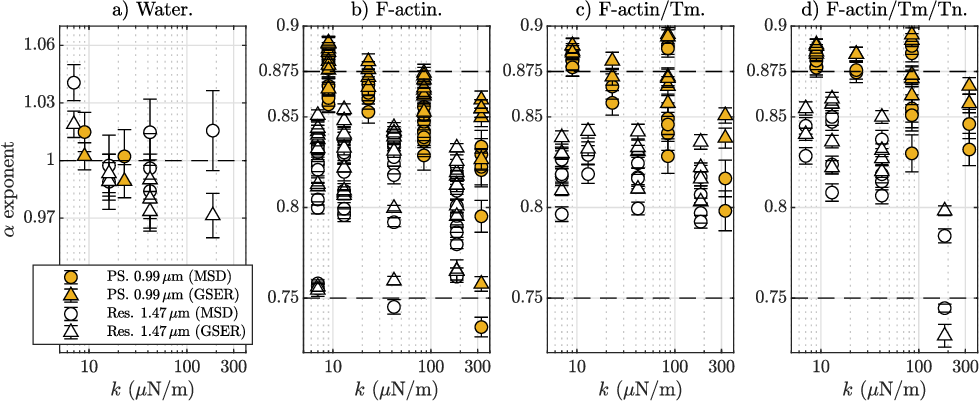}
\caption{\label{fig:supall} Values of the calculated power-law exponent $\alpha$ for each experiment (not averaged) in function of the trap stiffness, $k$. For clarity, we exclude the horizontal error bars and we limit the plots to the exponents obtained from the one dimensional particle mean-squared displacement (MSD) and the loss modulus ($G''$) calculated using the GSER. We show $\alpha$ values from the two type of particles: Polystyrene (PS) BSA $a=0.99\,\mu$m, from the MSD (\protect\tikzbullet{black}{orange}) and from $G''$ (\protect\tikztriangle{black}{orange}); melamine resin $a=1.47\,\mu$m, from the MSD (\protect\tikzbullet{black}{white}) and from $G''$ (\protect\tikztriangle{black}{white}). a) Measurements performed in water for control. b) F-actin. c) F-actin with tropomyosin (Tm). d) F-actin with Tm and troponin (Tn). The dashed lines (\protect\tikzdashedline{black} ) indicate the values $\alpha=7/8=0.875$ and $\alpha=3/4=0.75$.}
\end{figure*}

First, the bead radius, $a$, has to be larger enough than the mesh size, so that its motion reflects the mechanical properties of the surrounding medium. In our case, for $c_A = 1\,$mg/ml, the mesh size is $\xi = 0.3\,\mu$m, and the condition $a>\xi$ is verified for both tracer's sizes.
The persistence length, $l_p$, determines the classification of the entangled regime. For $c_A = 1\,$mg/ml, these semiflexible polymers verify $d \ll \xi < l_p \sim l$, where $d$ is the diameter of the proteins and $l$ is the average contour length. 
From a polymers-physics perspective, the length scale relevant for elastically active contacts in F-actin entangled networks is the entanglement length, $l_e$, which is the characteristic scale where an individual filament is sterically hindered because of other polymers nearby \cite{MacKintosh1995, Hinner1998, Gardel_microrheology_2003}. 
If $l_p\sim l$, the network of semiflexible polymers is considered to be in the \textit{highly entangled isotropic} regime, while if $\xi$ and $l_e$ are both much less than $l_p$, then the solution is in the \textit{tightly entangled} regime \cite{Tassieri2017}. 
For physiological values of F-actin, the experimental vale of the entanglement length is reported to be $l_e \sim 1-2\,\mu$m \cite{Isambert1996, palmer_diffusing_1999, addas_microrheology_2004}.
In the context of the tube model \cite{Isambert1996, Dichtl1999}, the entanglement length is related to the mesh size (tube width) and the persistence length by $l_e^3 = \xi^2\,l_p$. 
Using that expression, we obtain $l_e = 1.2\,\mu$m, a value in the local scale, i.e., very close to the size of the probes employed in our experiments.
The importance of the entanglement length, a relevant scale if we measure locally, on the order of microns, is confirmed by a constant value for the elastic modulus, $G'(\omega)$, over an extended range of frequencies, as shown in Fig. \ref{fig:sup9} and discussed in the Results Section.

Regarding this discussion on the characteristic length scales of the F-actin networks, it is possible to define a crossover length, in which the F-actin solution changes from bulk to intermediate mechanical behavior. The microrheological response at intermediate distances shows that the local viscoelasticity of the medium affects to the micro-mechanical response of the network at length scales larger than the expected characteristic lengths \cite{Sonn-Segev2014PRL}. 
According to the data of those studies \cite{Sonn-Segev2014SM}, we obtain that, in our case ($\xi/a=0.2$ and $0.3$), the lower limit for the crossover length is $r_0 \simeq 1.1\,\mu$m, matching the estimated value of the entanglement length. 
The PS beads here employed have a radius similar to those characteristic lengths, but resin probes may be big enough compared to the crossover length (25\%) to be able to detect the intermediate behavior between local and bulk mechanical properties. We observe this transition as an intermediate value of their power-law exponent in the form  of $3/4 < \alpha < 7/8$, 
as a consequence of the expected dependence of the micro-mechanical response of the fluid with the probe size in intermediate length scales.

Regarding the dependence of the power-law exponents with the applied optical trap, 
Fig. \ref{fig:fig3} shows a small decreasing of $\alpha$ with the applied optical trap for PS particles, 
which is not apparently observed for resin chemically-inert particles. 
BSA generates a less dense network around the PS tracers, producing a polymer-poor depletion zone because of the entropy loss of the molecules near the bead surface. In this case, the filamentous polymers are not attached to the bead, 
but surface adsorption may be already improved by the attractive optical force affecting the polymers surrounding the probe, 
causing a competition between the influence of BSA and the optical force. 
Experimentally, it has been observed that the increase of the depletion agent concentration in actin solutions decreases the exponent of the power-law behavior in the complex modulus \cite{Tharmann2006}. 

The different behaviors observed between the two types of probes are discernible by their VAFs 
which show different values for the power-law exponents $\beta$ (Fig. \ref{fig:sup12}).
The viscoelastic power-law behavior in the VAF of a microsphere in an actin solution was already observed by Xu \textit{et al.} \cite{Xu1998}. The value they calculated for the exponent in intermediate time-scales was $\beta = 0.97$, which is compatible with the value we obtain for chemically-inert resin particles. 
They also obtain the short and long times exponents: $1.95$ and $1.75$, respectively, very differentiated from the Newtonian $1.5$ and $3.5$, although a more rapid decay of the tail exponent at long times is expected because stress propagation is faster than diffusive in a viscoelastic fluid \cite{atakhorrami_short-time_2008}. 
Theoretically, Grebenkov and Vahabi \cite{Grebenkov2014} deduced explicit expressions for the statistical quantities considered here when the friction memory kernel decays as a power-law. 
Using their phenomenological model, which included inertial and hydrodynamic effects at short times and optical trapping at long times, they obtained power-law scaling in the VAFs at intermediate times for several $\alpha$ exponents. 
Using the data from the figures provided in their publication (Fig. 4 (c)), we obtain that the $\beta$ exponents are: $\beta \sim 0.98$ for $\alpha=7/8$, and  $\beta \sim 1.17$ for $\alpha=3/4$. 
A simple interpolation between these pairs of values provides $\beta=1.08$ for $\alpha=0.81$, which agrees with our experimental values for resin particles.
In any case, the exponent value $\beta \sim 0.98$ for $\alpha=7/8$ matches well with the experimental result obtained by Xu \textit{et al.}, indicating a actual value of $\alpha=7/8$ instead of the expected 3/4. 
However, the results of those studies confirm the atypical character of the exponent measured here for depleted particles, since the exponent $\beta < 1$ for depleted PS particles has not been previously observed. The physical interpretation of this exponent value is not clear, 
but the effects of the depletion layer are indeed more complex than expected, something confirmed by numerical studies of mean-field theory, which show that the depletion thickness and curvature effects depend on the interplay between the persistence and correlation polymer characteristic lengths \cite{Ganesan2008}. 

Finally, to further analyze the calculated $\alpha$ values and their dependency with the trap stiffness values, and the differences between simple F-actin networks and the F-actin/Tm/Tn complex, 
Fig. \ref{fig:supall} plots all the calculated subdiffusive exponents for the two types of probes. 
For clarity, it only includes $\alpha$ values obtained through the MSD and $G''$ (GSER). 
For control purpose, we include reference water measurements in Fig. \ref{fig:supall}a), 
returning values of $\alpha \approx 1$. 
In Fig. \ref{fig:fig3}, it can be seen that the F-actin/Tm/Tn complex generates values of $\alpha$ less influenced by the optical trap, at least for PS particles. 
This subtle effect can be clearly observed in the not-averaged data we plot in Fig. \ref{fig:supall}c) and d) in comparison with F-actin alone, Fig. \ref{fig:supall}b). 
Besides, the dependence of the $\alpha$ values with $k$ can be also observed in Fig. \ref{fig:supall}b), which indicates that the polymer network surrounding the probes is indeed being influenced by the external forces even for chemically-inert probes. 
Regarding the influence of Tm and Tn in the subdiffusive exponents, the values of $\alpha$ are closer to the 7/8 line when adding Tm, and even a little more when adding Tm and Tn, and this effect appears for both kinds of particles. 
In other words, the subdiffusive exponents are less dispersed when using Tm or Tm/Tn. 
The effect of the increasing of filament stiffness by a 50\% has been only observed in rheological measurements at very low frequencies  appearing for elastic and loss modulus smaller than $0.1\,$Pa (molar ratio of 7:1:1, same as here), 
whereas at frequencies greater than $10\,$Hz, the differences in the complex modulus because of the presence or absence of Tm/Tn are apparently negligible \cite{Goldmann2000}.     
Our results show that the F-actin/Tm/Tn complex is able to slightly modify the mechanical response of the filaments in the local microscale, something that can be understood as an improvement in the stabilization of the filaments by adding Tm and Tn.

\section{Conclusions}

In this work, we investigate the micro-mechanical behavior of F-actin solutions, with focus on the skeletal thin filament, through their power-law behavior when analyzing the local fluctuations generated by two different kind of microtracers: chemically-inert melamine resin and polystyrene depleted particles. We use optical trapping interferometry to measure, with nanometric resolution in the microsecond range, the Brownian position fluctuations of these single optically-trapped microbeads. 
The short-time/high-frequency regime permits to study, in a range of several decades, the power-law behavior of the motion of the trapped microtracers immersed in F-actin networks, reflecting their interaction with the individual filaments.
When we use one-particle microrheology, the obtained results depend on the characteristic lengths of the network in relation with the size of the tracer bead, i.e., this technique allows to investigate the mechanical response on the biological scale conditioned to a correct characterization of the involved length scales. 
We have observed that the values of the local subdiffusive exponent ($\alpha \simeq 7/8$), obtained from the MSD$(t)$, PSD$(t)$ and $G''(\omega)$, depend on the bead's properties and its interaction with the surrounding polymer. 
We have determined that the characteristic length scale for these networks at a concentration of $1\,$mg/ml is $\sim 1.2\,\mu$m. Our resin tracers are slightly bigger ($1.5\,\mu$m) and they may detect the transition between local and bulk microrheology, and thus providing an experimental $\alpha$ value intermediate between 7/8 and 3/4. 
These exponents obtained from chemically inert tracers are not observed to depend on the applied external optical forces, but the ones from the depleted PS (BSA-added) particles do.   
This difference of behavior is confirmed by the VAFs, by showing a non-expected value of its power-law exponent for depleted PS particles ($\beta < 1$), without clear dependencies on other parameters.  

Finally, through this experimental system, we have studied the micro-mechanical response of the skeletal thin filament. The F-actin/Tm/Tn complex has a role in muscle contraction through its interaction with myosin, and it is expected to be stiffer than F-actin alone.
We have not detected an increase of stiffness when adding skeletal tropomyosin and troponin to F-actin, because our experimental system does not detect the very low-frequency elastic plateau, and we also have to take into account the limitations induced by one-particle microrheology and the influence of the optical elastic force in the fluid itself. 
The plotting of all the results for the power-law exponent values in Figure \ref{fig:supall} shows how the dispersion around the 7/8 value is reduced when including Tm or Tm/Tn to the F-actin network, effect that we interpret as a micro-mechanical stabilization of the filaments.  
The results here contained can be of great utility for expanding the knowledge of several related fields: for example, to indicate further directions in the development of new theoretical aspects in the colloids-polymer interactions, with potential application in bioengineering.

\section*{Conflict of interest}

There are no conflicts to declare.

\section*{Acknowledgments}
P.D.G acknowledges support aid by grant PID2020-117080RB-C54 funded by MCIN/AEI/10.13039/501100011033, and J. C. G\'omez-S\'aez for her proofreading of the texts. J.R.P. acknowledges support from National Institutes of Health grant R01 HL128683. A.A. acknowledges funding from the Swiss National Science Foundation through project PP00P2\_202661. We acknowledge the fruitful commentaries provided by M. Tassieri.

\providecommand*{\mcitethebibliography}{\thebibliography}
\csname @ifundefined\endcsname{endmcitethebibliography}
{\let\endmcitethebibliography\endthebibliography}{}

\end{document}